# Simultaneous Extraction of Charge Density Dependent Mobility and Variable Contact Resistance from Thin Film Transistors.

Riccardo Di Pietro[1]*, Deepak Venkateshvaran[2], Andreas Klug[3], Emil J.W. List-Kratochvil[3,4], Antonio Facchetti[5], Henning Sirringhaus[2] and Dieter Neher[1]

[1]*Universität Potsdam, Institut für Physik und Astronomie, Karl-Liebknecht-Str. 24-25, 14476 Potsdam, Germany*

[2]*Cambridge University, Cavendish Laboratory, J. J. Thomson Avenue, CB3 0HE Cambridge, United Kingdom*

[3]*NanoTecCenter Weiz Forschungsgesellschaft mbH, Franz-Pichler-Straße 32, A-8160 Weiz, Austria*

[4] *Institute of Solid State Physics, Graz University of Technology, Petersgasse 32, A-8010 Graz, Austria*

[5] *Polyera Corporation, 8045 Lamon Ave, STE 140, Skokie, IL 60077-5318, USA*

\* dipietro@uni-potsdam.de

A model for the extraction of the charge density dependent mobility and variable contact resistance in thin film transistors is proposed by performing a full derivation of the current-voltage characteristics both in the linear and saturation regime of operation. The calculated values are validated against the ones obtained from direct experimental methods. This approach allows unambiguous determination of both contact and channel resistance from the analysis of the current voltage characteristics of a single device, with no a-priori assumption on the two parameters. It solves the inconsistencies in the commonly accepted mobility extraction methods and provides new possibilities for the analysis of the injection and transport processes in semiconducting materials.

Field-effect transistors (FETs) are one of the most widely used tools for the analysis of the charge transport properties of semiconducting materials [1,2]. While the working principle is well understood and characterized [3], there is still no general analytical solution for the current voltage characteristics in presence of charge density dependent mobility [4], but only limited solutions for a specific functional dependence of mobility on gate voltage [5]. The different effects of non-constant mobility in the linear and saturation regime of operation usually lead to different values for mobility extracted in the different regimes [6]. To complicate matters even further, mobility extraction is also hindered by the presence of contact resistance, which is as well dependent on the regime of operation of the transistor [7,8] and leads to artifacts in the mobility extraction procedure.

In order to disentangle the effects of charge density dependent mobility and contact resistance, several techniques have been developed [9], from adaptations of the transmission line method (TLM) [10–16] to more complex techniques such as four point probe (FPP) measurements [17–19] and scanning Kelvin probe microscopy (SKPM) [20,21]. Although these techniques allow for a direct measurement of the two parameters, they either involve the comparison of multiple devices (TLM), or are limited to a specific device structure (FPP and SKPM), thus restricting their applicability only to a subset of possible device architectures. To bridge this gap, several analytical models have been proposed to extract the effects of contact resistance from the analysis of the electrical characteristics of working devices [5,6,22–24]. However, the analytical description is complicated by the fact that both mobility and contact resistance are usually gate voltage dependent, and all the method proposed so far for mobility extraction either assume a constant contact resistance or they still require the comparison of devices with different channel lengths in order to separate the influence of the gate voltage on the two parameters. The method proposed by Reese and Bao allows extracting the two parameters but is however only applicable for a contact resistance which does not depend on the source-drain voltage [6].

In this letter we report a derivation of the standard textbook current-voltage equation for field-effect transistors [3] to include, ab-initio, the effects of charge density dependent mobility $\mu(V)$ and variable contact resistance, without the need to assume any a priori functional dependence of the two parameters on the applied voltages. The explicit inclusion of $\mu(V)$ introduces an additional correction factor in the extraction of the mobility in the saturation regime. A simple method to take into account this correction is proposed which allows to accurately estimate both charge density dependent mobility and variable contact resistance for any given device geometry from the analysis of the transfer characteristics only. By comparing the obtained values with experimental results obtained from gated FPP and gated TLM measurements, one finds that in all cases the calculated contact resistance and mobility agree both in magnitude and in charge density dependence with the experimental results. The gained flexibility (only the standard saturation and linear transfer characteristics need to be acquired) is demonstrated in the analysis of the influence of organic semiconductor thickness and processing conditions on contact resistivity.

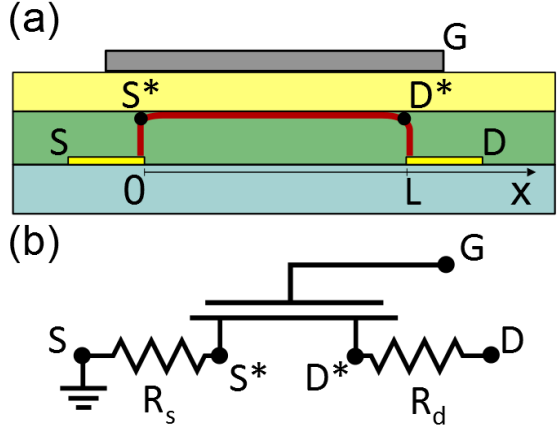

Figure 1. (a) Schematic cross-section of a bottom-contact top-gate transistor, showing the main parameters used in the mathematical derivation. (b) Equivalent circuit of the transistor used to include the effect of source ($R_S$) and drain ($R_D$) resistance ($V_G$ dependent), with the different points at which the potential is evaluated in the text.

Figure 1 shows a schematic of a field-effect transistor (a) with the equivalent circuit and all the relevant parameters used in the text (b). The transistor is operated in a common source configuration $(V_S = 0V)$. In the following derivation each potential is written either as $V_X$, the potential at position $X$ with respect to the ground, or $V_{XY}$, the potential difference between two points $X$ and $Y$, with the exception of $V(x)$ which represents the potential along the channel referred to ground. The full mathematical derivation is reported in the supplementary information.

The charge density per unit area in the channel is defined using the differential capacitance per unit area $C(V)$:

$$e \cdot n(V_G - V(x)) = \int_0^{V_G - V(x)} C(v) dv \qquad (1)$$

For generality's sake $C(V)$ is kept implicit during the derivation, its value can be easily obtained either by calculating the geometric capacitance of the dielectric layer, or by directly performing a capacitance-voltage measurement. The field-effect mobility is defined as the average mobility of all charge accumulated at point $x$ for a given voltage $V_G - V(x)$:

$$\langle \mu(V_G - V(x)) \rangle = \frac{\int_0^{V_G - V(x)} \mu(v) C(v) dv}{\int_0^{V_G - V(x)} C(v) dv} \qquad (2)$$

$\mu(V)$, the charge density dependent mobility, represents the mobility of the additional charge that is introduced in the channel when the gate voltage is increased from $V$ to $V + dV$. The channel conductivity $\sigma = en\langle\mu\rangle$ can then be written as:

$$\sigma(V_G - V(x)) = \int_0^{V_G - V(x)} \mu(v) C(v) dv \qquad (3)$$

Using equation (3) in the standard textbook derivation for the gradual channel approximation [3] the main field-effect transistor equation is obtained:

$$IL = W \int_{V_{S*}}^{V_{D*}} \sigma(V_G - V(x)) dV(x) \qquad (4)$$

W and L are the length and width of the transistor's channel and $V_{S*}$ and $V_{D*}$ are unknown. In order to evaluate the integral in equation (4) without any restriction on the unknown parameters, $\sigma(V)$ is replaced with its Taylor expansion at $V = V_G$:

$$\sigma(V_G - V(x)) = \sum_{i=0}^{\infty} \frac{\sigma^{(i)}(V_G)}{i!}(-V(x))^i \qquad (5)$$

Mobility can be extracted combining equation (4) and (5) and calculating the first derivative of the current against the gate voltage:

$$\frac{dI}{dV_G} \cong \frac{W}{L} \mu(V_G) C(V_G) V_{D*S*} \qquad (6)$$

where all the terms containing $\frac{dV_{D*S*}}{dV_G}$ are neglected, and the series expansion can be limited to $i = 1$ (since in the linear regime $V_D \ll V_G$, the argument in equation (5) is always close to $V_G$). Using $V_{DS}$ instead of the effective channel voltage $V_{D*S*}$ leads to an underestimation of the field-effect mobility in presence of contact resistance.

In the saturation regime it is safe to assume $V_{D*} = V_G$ given the higher channel resistance due to the channel being pinched-off close to the drain electrode. Starting from equation (4), $\left(\frac{d\sqrt{I}}{dV_G}\right)^2$ is evaluated using again the Taylor expansion in equation (5). In order to perform the derivation it is assumed that:

$$\frac{dV_{S*}}{dV_G} \ll 1 \qquad (7)$$

SKPM measurements performed in the saturation regime [25,26] show that in this case the voltage drop at the source contact is usually small compared to the applied gate voltage, so that condition (7) is not so restrictive. With this approximation, it is possible to obtain the following equation, where all the higher order terms ($i > 1$) of the Taylor series are grouped in a generic correction factor $k(V_G)$:

$$\left(\frac{d\sqrt{I}}{dV_G}\right)^2 = \frac{W}{2L} \mu(V_G) C(V_G) \cdot k(V_G) \qquad (8)$$

The correction factor depends on the particular shape of $\mu(V_G)$. While in general the value of $k$ is gate voltage dependent, it is easy to show how for $\mu(V_G)C(V_G) = C_i \mu_0 (V_G - V_{th})^\beta$ (power laws are usually employed to approximate charge density dependent mobility in OFETs [5,24]) equation (8) becomes:

$$\left(\frac{d\sqrt{I}}{dV_G}\right)^2 = \frac{W}{2L} C_i \mu_0 (V_G - V_{th})^\beta \frac{(\beta+2)}{2(\beta+1)} \quad (9)$$

and the correction factor becomes gate voltage independent and equal to:

$$k = \frac{\beta+2}{2(\beta+1)} \quad (10)$$

For positive values of $\beta$ (mobility increasing with charge density [4,27]), $k$ will be limited between 1 ($\beta = 0$) and 0.5 ($\beta \to \infty$). As it is always possible to approximate the gate voltage dependence of the mobility in the form of a polynomial function, for any monotonically increasing function the following inequality holds true:

$$0.5 \leq k(V_G) \leq 1 \; \forall V_G \quad (11)$$

It is therefore possible to obtain an average value of $k$ by simply fitting $\left(\frac{d\sqrt{I}}{dV_G}\right)^2$ with a power law function and using the extracted value of $\beta$ to calculate $k$ (figure S1 and S2, supplementary information). Although a power law is used to extract the correction factor, this method gives the possibility to extract an approximate value of the gate voltage dependent mobility without forcing any a priori explicit functional dependence. The consistency of the results is shown by the agreement with the second derivative method (figure S1, supplementary information). However, equation (8) produces far less noise in the result due to the lower order derivation, allowing for faster measurements. The standard saturation mobility equation, which neglects this correction factor will lead to an underestimation of the calculated mobility value [28].

It is noteworthy to stress that under the above mentioned conditions, the quantity $\mu(V_G)C(V_G)$ appearing in equations (6) and (8) is exactly the same one. It is therefore straightforward to obtain the channel voltage $V_{D*S*}$, free of any contact resistance contribution, from the ratio of linear and saturation mobility:

$$\frac{\frac{dI_{lin}}{dV_G} \cdot k}{2\left(\frac{d\sqrt{I_{sat}}}{dV_G}\right)^2} \cong V_{D*S*}(V_G) \quad (12)$$

Knowing the voltage drop along the channel, it is possible to estimate the total contact resistance $R_C = R_S + R_D$ by simply dividing the voltage drop at the contacts by the measured current, without assuming any specific dependence on the gate voltage:

$$R_C(V_G, V_{DS}) = \frac{(V_{DS} - V_{D^*S^*}(V_G))}{I_d^{lin}} \quad (13)$$

where the dependence of $R_C$ on $V_G$ and $V_{DS}$ is evident. The possibility of extracting $R_C(V_G, V_{DS})$ and charge density dependent $\mu(V_G)$ for any transistor geometry and without requiring any additional measurement makes this approach extremely versatile. Equation (7) is the only extra assumption required in order for the model to provide consistent results.

The validity of this model is now assessed by preparing FPP field-effect transistors and a set of different channel length transistors for gated-TLM characterization, where experimental and analytical extraction methods can be compared on the very same device using the same set of measurements. All devices were prepared on photolithographically defined gold source-drain electrodes on glass substrates, using either Poly[2,5-bis(3-tetradecylthiophen-2-yl)thieno[3,2-b]thiophene] (pBTTT) [29] or poly{[N,N'-bis(2-octyldodecyl)-naphthalene-1,4,5,8-bis(dicarboximide)-2,6-diyl]-alt-5,5'-(2,2'-bithiophene)} (P(NDI2OD-T2)) [30] as semiconducting layer, poly-methyl-methacrylate (PMMA, $\varepsilon_r = 3.6$) as dielectric layer and aluminum as gate electrode (full details on device preparation and geometry are available in the supplementary information).

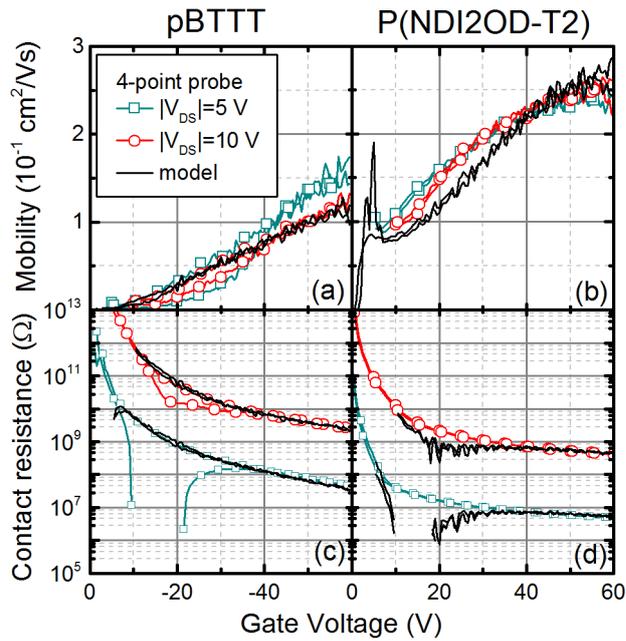

Figure 2. Comparison of the mobility extracted from the four point probe measurement (lines + symbols ) for $|V_{DS}|=5V$ and $|V_{DS}|=10V$, positive for P(NDI2OD-T2) and negative for pBTTT and the charge density dependent mobility extracted using equation (8) (black line) for p-type (a) (pBTTT) and n-type (b) (P(NDI2OD-T2). (c) and (d) Comparison of the contact resistance extracted with the four point probe (lines + symbols) and equation (12) (black line), for pBTTT and P(NDI2OD-T2) respectively. Curves for $|V_{DS}|=10V$ are offset by two orders of magnitude for clarity.

In figure 2(a) and 2(b), the charge density dependent mobility extracted from the FPP measurement in the linear regime (using the linear extrapolation of the channel potential [17]) is compared with the mobility extracted from the saturation transfer characteristics using equation (8). The displayed results have been obtained by estimating the differential capacitance from the geometric one. In the case of pBTTT, the agreement between the two methods is particularly good for $V_{DS} = -10V$, while the experimental curve for $V_{DS} = -5V$ suffers from a slightly offset voltage measured on the voltage probe close to the source. The agreement is slightly less good for P(NDI2OD-T2), with the mobility extracted from equation (8) underestimating the value obtained from the FPP measurements at low gate voltages.

In figure 2(c) and (d), the extracted value for the contact resistance is shown both for the FPP measurements and using equation (13), the latter demonstrating how the analytical method recovers the full dependence of $R_C$ on both $V_{DS}$ and $V_{GS}$. For the PBTTT device the agreement is very good through the whole range (the traces overlap for most of the graphs), with the exception of the experimental measurement at $V_{DS} = -5V$ which again suffers from the offset on the voltage measured on the source probe (hence the drop in the measured contact resistance). Also for P(NDI2OD-T2) (figure 2(d)) the agreement with the experimental data is particularly good at high gate voltages ($V_G \geq 20V$). At low gate voltages the saturation mobility becomes equal or lower than the linear one (figure S3, supplementary information) leading to a negative contact resistance, which might be related to the condition in equation (7) not being met for this device at low gate voltages, probably due to some degradation occurring during the patterning process (the P(NDI2OD-T2) devices used for TLM characterization do not show this effect).

The main limiting factor determining the uncertainty in the calculated parameters is due to the accuracy of the measurement setup (the relative error on current measurement for the measurement setup is $\xi_i \cong 10^{-3}$), since the uncertainty in the estimation of channel dimensions and capacitance will have the same impact on calculated and measured values, canceling out when comparing the results. Taking into account the numerical derivation, the final accuracy $\xi_C$ is usually between 3 and 10 % for both mobility and contact resistance (a detailed analysis of the uncertainty for the FPP measurement is reported in the supplementary information, figure S4). While at low voltages the differences between measured and calculated values are significant for the reasons explained in the previous paragraph, at high gate voltages the difference between calculated and measured parameters is always lower than $2 \cdot \xi_C$, with the proposed method extracting values that are in full agreement with the ones obtained by direct experimental methods.

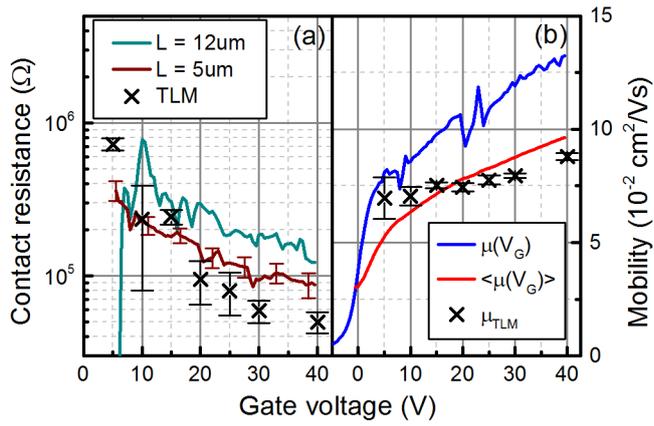

Figure 3. (a) Comparison of the contact resistance of P(NDI2OD-T2) transistors extracted with the TLM method and with equation (13) for the transistors with channel length of 5 μm and 12 μm. Error bars, shown only for the 5 μm channel for clarity, are similar for the 12 μm channel length. (b) Comparison of the mobility extracted from the slope of the TLM plots and the charge density dependent and average field-effect mobility calculated using the analytical model for the 5 μm channel transistor.

We performed a similar comparison with the TLM method [31]. The total resistance for a set of P(NDI2OD-T2) transistors against different channel lengths, extracted for different gate voltages from the slope of the output curve between $V_{DS} = 1V$ and $V_{DS} = 4V$ (figure S5 supplementary information), has been fitted with a straight line according to the standard TLM equation (adapted for a gate voltage dependent mobility) [11]:

$$R_D = R_C + \frac{L}{W\langle\mu\rangle ne} = R_C + \frac{L}{W\int_0^{V_{GS^*}} \mu(v)C(v)dv} \tag{14}$$

Figures 3(a) and 3(b) show contact resistance and charge density dependent mobility respectively, extracted from the TLM measurements and using the analytical model for devices with 5 µm and 12 µm channel length. As can be seen the contact resistance extracted from the analytical method has the same gate voltage dependence as the values extracted from the TLM analysis. The different gate voltage dependence of contact resistance compared to the 4 point probe devices (figure 2) is due to the much larger overlap between the gate and source-drain electrodes in the TLM devices, leading to a much longer effective contact length. The contact resistance extracted for the 5 µm channel device is slightly lower than for the 12 µm one, which is probably due to device to device variation. In figure 3(b) the mobility extracted by the two methods are compared, showing how the mobility extracted from the TLM analysis correctly matches with the average mobility (calculated using equation (2)). The discrepancy usually observed between the mobility values extracted from the transfer characteristics and from the output curves is shown here to be due to the physically different parameter measured in the two cases, and not to an approximation of the analytical model [6].

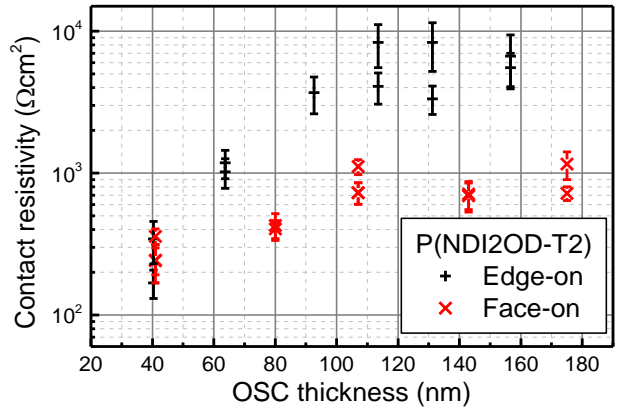

Figure 4. Contact resistivity at $V_G = 58V$ for P(NDI2OD-T2) transistors with different thicknesses of the OSC layer, showing the influence of the increased bulk resistance on contact resistivity. The difference in bulk conductivity between face-on and edge-on polymer chains is lost for thinner films, suggesting an altogether different morphology on top of the gold source drain electrodes.

In figure 4, the developed model is used to study the dependence of contact resistivity on the thickness of the semiconducting layer (at $V_G = 58V$) for two sets of P(NDI2OD-T2) transistors with different preparation conditions (a scenario where TLM measurement would be unpractical for the high number of devices required and four point probe would be complicated by the thickness of some of the measured devices). We compared a set of devices prepared using chlorobenzene (CB), a solvent that promotes strong aggregation in the polymer film, and produces films with high electron mobility (~0.2 cm$^2$/Vs) and a face-on orientation of the polymer chains with respect to the substrate, with a set of devices prepared using a 1:1 mixture of chloronaphthalene:xylene (CN:Xyl), which inhibits aggregation, leading to films with lower charge carrier mobility (~0.06 cm$^2$/Vs) and edge-on orientation of the polymer chains [32,33]. Contact resistivity was obtained from the contact resistance values using the current crowding model proposed by Chiang et al. [15]. As expected from the different device preparation conditions, for large thicknesses there is a clear difference in contact resistivity due to the higher bulk resistivity of the CN:Xyl films. However, for film thicknesses of roughly 40 nm contact resistance of both films becomes comparable, therefore suggesting an altogether different film morphology on top of the gold source and drain electrodes which extends for more than 10 nm in the bulk. The origin of this behavior is currently under investigation.

In conclusion, a very simple and compact model is proposed for the extraction of charge density dependent mobility and variable contact resistance which can be easily applied to any transistor as it does not require a particular device geometry. It allows for the first time to extract accurate values for both charge density dependent mobility and contact resistance, completely separating the contribution of each parameter to the total current without any a-priori assumption or the analysis of devices with multiple channel lengths. By taking into account ab initio both contact resistance and charge density dependent mobility it finally solves all the inconsistencies between the mobility extracted from the linear and saturation regime of the transfer characteristics and from the output curves, providing an accurate estimation of the carrier mobility. As such, this model can be a valuable tool to characterize charge-transport mechanisms and charge-carrier injection in field-effect transistors, allowing for a significant improvement of the current understanding of the charge-transfer processes at the metal-organic semiconductor interface.

This work was financially supported by the Deutsche Forschungsgemeinschaft (DFG) within the Collaborative Research Centre HIOS (SFB 951). A. K. and E. L.-K. gratefully acknowledge the financial support of the Styrian Government (project BioOFET 2, GZ:A3-11.B-36/2010-5). R. D. P. would like to thank Mr. Daniel Pinkal for useful discussion during the preparation of the manuscript.

[1]    R. A. Street, Adv. Mater. **21**, 2007 (2009).

[2]    C. D. Dimitrakopoulos and D. J. Mascaro, IBM J. Res. Dev. **45**, 11 (2001).

[3]    S. M. Sze, *Semiconductor Devices. Physics and Technology* (WILEY, 1985), p. 523.

[4]    M. C. J. M. Vissenberg and M. Matters, Phys. Rev. B **57**, 12964 (1998).

[5]    G. Horowitz, P. Lang, M. Mottaghi, and H. Aubin, Adv. Funct. Mater. **14**, 1069 (2004).

[6]    C. Reese and Z. Bao, J. Appl. Phys. **105**, 024506 (2009).

[7]    M. Gruber, E. Zojer, F. Schürrer, and K. Zojer, Adv. Funct. Mater. **23**, 2941 (2013).

[8]    B. Hamadani, H. Ding, Y. Gao, and D. Natelson, Phys. Rev. B **72**, 235302 (2005).

[9]    D. Natali and M. Caironi, Adv. Mater. **24**, 1357 (2012).

[10]   G. K. Reeves and H. B. Harrison, IEEE Electron Device Lett. **EDL-3**, 111 (1982).

[11]   Y. Xu, R. Gwoziecki, I. Chartier, R. Coppard, F. Balestra, and G. Ghibaudo, Appl. Phys. Lett. **97**, 063302 (2010).

[12]   J. Zaumseil, K. W. Baldwin, and J. A. Rogers, J. Appl. Phys. **93**, 6117 (2003).

[13]   E. J. Meijer, G. H. Gelinck, E. van Veenendaal, B.-H. Huisman, D. M. de Leeuw, and T. M. Klapwijk, Appl. Phys. Lett. **82**, 4576 (2003).

[14]   H. Sirringhaus, N. Tessler, D. S. Thomas, P. J. Brown, and R. H. Friend, Adv. Solid State Phys. **39**, 101 (1999).

[15]   C. Chiang, S. Martin, J. Kanicki, Y. Ugai, T. Yukawa, and S. Takeuchi, Jpn. J. Appl. Phys. **37**, 5914 (1998).

[16]   S. Gamerith, A. Klug, H. Scheiber, U. Scherf, E. Moderegger, and E. J. W. List, Adv. Funct. Mater. **17**, 3111 (2007).

[17]   P. V. Pesavento, K. P. Puntambekar, C. D. Frisbie, J. C. McKeen, and P. P. Ruden, J. Appl. Phys. **99**, 094504 (2006).

[18]   T. J. Richards and H. Sirringhaus, J. Appl. Phys. **102**, 094510 (2007).

[19]   M. Caironi, M. Bird, D. Fazzi, Z. Chen, R. Di Pietro, C. Newman, A. Facchetti, and H. Sirringhaus, Adv. Funct. Mater. **21**, 3371 (2011).

[20]   K. Seshadri and C. D. Frisbie, Appl. Phys. Lett. **78**, 993 (2001).

[21]   L. Bürgi, T. J. Richards, R. H. Friend, and H. Sirringhaus, J. Appl. Phys. **94**, 6129 (2003).

[22]   R. A. Street and A. Salleo, Appl. Phys. Lett. **81**, 2887 (2002).

[23]   P. V. Necliudov, M. S. Shur, D. J. Gundlach, and T. N. Jackson, J. Appl. Phys. **88**, 6594 (2000).

[24]   D. Natali, L. Fumagalli, and M. Sampietro, J. Appl. Phys. **101**, 014501 (2007).

[25]   G. Lu, J. Blakesley, S. Himmelberger, P. Pingel, J. Frisch, I. Lieberwirth, I. Salzmann, M. Oehzelt, R. Di Pietro, A. Salleo, N. Koch, and D. Neher, Nat. Commun. **4**, 1588 (2013).

[26]   L. Bürgi, T. Richards, M. Chiesa, R. H. Friend, and H. Sirringhaus, Synth. Met. **146**, 297 (2004).

[27]   H. Bässler and A. Köhler, Top. Curr. Chem. **312**, 1 (2012).

[28]   H. Klauk, Chem. Soc. Rev. **39**, 2643 (2010).

[29]   I. McCulloch, M. Heeney, C. Bailey, K. Genevicius, I. Macdonald, M. Shkunov, D. Sparrowe, S. Tierney, R. Wagner, W. Zhang, M. L. Chabinyc, R. J. Kline, M. D. McGehee, and M. F. Toney, Nat. Mater. **5**, 328 (2006).

[30]   H. Yan, Z. Chen, Y. Zheng, C. Newman, J. R. Quinn, F. Dötz, M. Kastler, and A. Facchetti, Nature **457**, 679 (2009).

[31]   A. Klug, A. Meingast, G. Wurzinger, A. Blümel, K. Schmoltner, U. Scherf, and E. J. W. List, Proc. SPIE **8118**, 811809 (2011).

[32]   R. Steyrleuthner, R. Di Pietro, B. A. Collins, F. Polzer, S. Himmelberger, M. Schubert, Z. Chen, S. Zhang, A. Salleo, H. Ade, A. Facchetti, and D. Neher, J. Am. Chem. Soc. **Submitted**, (2013).

[33]   S. Fabiano, H. Yoshida, Z. Chen, A. Facchetti, and M. A. Loi, ACS Appl. Mater. Interfaces **5**, 4417 (2013).

# Simultaneous Extraction of Charge Density Dependent Mobility and Variable Contact Resistance from Thin Film Transistors

**Supporting information**

*Riccardo Di Pietro[1]\*, Deepak Venkateshvaran[2], Andreas Klug[3], Emil J.W. List-Kratochvil[3,4], Antonio Facchetti[5], Henning Sirringhaus[2] and Dieter Neher[1]*

[1]*Universität Potsdam, Institut für Physik und Astronomie, Karl-Liebknecht-Str. 24-25, 14476 Potsdam, Germany*

[2]*Cambridge University, Cavendish Laboratory, J. J. Thomson Avenue, CB3 0HE Cambridge, United Kingdom*

[3]*NanoTecCenter Weiz Forschungsgesellschaft mbH, Franz-Pichler-Straße 32, A-8160 Weiz, Austria*

[4] *Institute of Solid State Physics, Graz University of Technology, Petersgasse 32, A-8010 Graz, Austria*

[5] *Polyera Corporation, 8045 Lamon Ave, STE 140, Skokie, IL 60077-5318, USA*

*\* dipietro@uni-potsdam.de*

## 1 - Full derivation of the linear mobility equation

As reported in the main text, the amount of charge in the channel of a transistor is defined through the differential capacitance per unit area $C(V) = \frac{dq(V)}{dV}$, while the field-effect mobility in the drift current equation is defined as the average value of mobility over the charge density.

$$ne = \int_0^{V_G - V(x)} C(v) dv \qquad (1)$$

$$\langle \mu(V_G - V(x)) \rangle = \frac{\int_0^{V_G - V(x)} \mu(v) C(v) dv}{\int_0^{V_G - V(x)} C(v) dv} \qquad (2)$$

$\mu(V)$, the charge density dependent mobility, in this context represents the mobility of the charge that is added to the system when the gate voltage is increased from $V$ to $V + dV$.

Using equations (1) and (2) in the gradual channel approximation, the drift current equation can be written as:

$$I(x) = W \cdot ne(V_G - V(x)) \cdot \langle \mu(V_G - V(x)) \rangle \frac{dV(x)}{dx} = W \left( \int_0^{V_G - V(x)} \mu(v) C(v) dv \right) \frac{dV(x)}{dx} \qquad (3)$$

By separating the variables in equation (3) it is possible to write the master equation for charge transport in field-effect transistors:

$$I = \frac{W}{L} \int_{V_{S^*}}^{V_{D^*}} \left( \int_0^{V_G - V(x)} \mu(v) C(v) dv \right) dV(x) \qquad (4)$$

In the linear regime $V_{S^*}, V_{D^*} \ll V_G$. In order to extract the mobility from equation (4) channel conductivity $\sigma(V_G - V(x))$ is considered:

$$ne(V_G - V(x)) \cdot \langle\mu(V_G - V(x))\rangle = \sigma(V_G - V(x)) = \int_0^{V_G-V(x)} \mu(v)C(v)dv \qquad (5)$$

and $\sigma(V_G - V(x))$ is expanded around $V_G$ in a Taylor series:

$$\sigma(V_G - V(x)) = \sum_{i=0}^{\infty} \frac{\sigma^{(i)}(V_G)}{i!}(-V(x))^i \qquad (6)$$

From which it is possible to write the generalized integral of $\sigma(V_G - V(x))$ as:

$$\int \sigma(V_G - V(x))dV(x) = -\sum_{i=0}^{\infty} \frac{\sigma^{(i)}(V_G)}{(i+1)!}(-V(x))^{i+1} \qquad (7)$$

Equation (4) can therefore be rewritten as:

$$I = \frac{W}{L}\left[-\sum_{i=0}^{\infty} \frac{\sigma^{(i)}(V_G)}{(i+1)!}(-V(x))^{i+1}\right]_{V_{S^*}}^{V_{D^*}} \qquad (8)$$

And now the derivative with respect to the gate voltage can be easily calculated, and if the series is limited to the first order (which is not a strong assumption since $V_{S^*}, V_{D^*} \ll V_G$, therefore the Taylor series is always evaluated close to $V_G$), and neglect terms containing $\frac{dV_{D^*S^*}}{dV_G}$ for the same reasons, the equation for the mobility extraction in the linear regime can be written as:

$$\frac{dI}{dV_G} \cong \frac{W}{L}\mu(V_G)C(V_G)V_{D^*S^*} \qquad (9)$$

## 2 - Full derivation of the saturation mobility equation

In the saturation regime it is safe to assume $V_{D^*} = V_G$, therefore equation (4) can be written as:

$$I = \frac{W}{L}\int_{V_{S^*}}^{V_G}\left(\int_0^{V_G-V(x)} \mu(v)C(v)dv\right)dV(x) \qquad (10)$$

The simplest way to obtain $\mu(V)C(V)$ is to perform a second derivative of the saturation current:

$$\frac{d^2I}{dV_G^2} = \frac{W}{L}\mu(V)C(V) \qquad (11)$$

However, performing a second derivative on a transfer curve increases introduces a high noise in the result, thus requiring longer integration times and/or more data points to obtain a reasonable signal to noise ratio.

In order to perform only a single derivation also in the saturation regime, it is sufficient to calculate the following derivative:

$$\left(\frac{d\sqrt{I}}{dV_G}\right)^2 = \left(\frac{1}{2\sqrt{I}}\frac{dI}{dV_G}\right)^2 = \frac{1}{4|I|}\left(\frac{dI}{dV_G}\right)^2 \qquad (12)$$

Replacing the integral in equation (10) with equation (7), the first derivative can be expressed as:

$$\frac{dI}{dV_G} = \frac{d}{dV_G}\left\{\frac{W}{L}\left[-\sum_{i=0}^{\infty}\frac{\sigma^{(i)}(V_G)}{(i+1)!}(-V(x))^{i+1}\right]_{V_{S^*}}^{V_G}\right\} =$$

$$\frac{W}{L}\frac{d}{dV_G}\left[-\sum_{i=0}^{\infty}\frac{\sigma^{(i)}(V_G)}{(i+1)!}(-V_G)^{i+1} + \sum_{i=0}^{\infty}\frac{\sigma^{(i)}(V_G)}{(i+1)!}(-V_{S^*})^{i+1}\right] = \frac{W}{L}\left[-\sum_{i=1}^{\infty}\frac{\sigma^{(i)}(V_G)}{(i)!}(-V_G)^i + \right.$$

$$\left.\sum_{i=0}^{\infty}\frac{\sigma^{(i)}(V_G)}{(i)!}(-V_G)^i + \sum_{i=1}^{\infty}\frac{\sigma^{(i)}(V_G)}{(i)!}(-V_{S^*})^i - \sum_{i=0}^{\infty}\frac{\sigma^{(i)}(V_G)}{(i)!}(-V_{S^*})^i\frac{dV_{S^*}}{dV_G}\right] =$$

$$\frac{W}{L}\left[\sigma(V_G) + \sum_{i=1}^{\infty}\frac{\sigma^{(i)}(V_G)}{(i)!}(-V_{S^*})^i - \sum_{i=0}^{\infty}\frac{\sigma^{(i)}(V_G)}{(i)!}(-V_{S^*})^i\frac{dV_{S^*}}{dV_G}\right] \tag{13}$$

While the current can be written as:

$$I = \frac{W}{L}\int_{V_{S^*}}^{V_G}\sigma(V_G - V(x))dV(x) = \frac{W}{L}\left[-\sum_{i=0}^{\infty}\frac{\sigma^{(i)}(V_G)}{(i+1)!}(-V(x))^{i+1}\right]_{V_{S^*}}^{V_G} =$$

$$\frac{W}{L}\left[-\sum_{i=0}^{\infty}\frac{\sigma^{(i)}(V_G)}{(i+1)!}\left((-V_G)^{i+1} - (-V_{S^*})^{i+1}\right)\right] \tag{14}$$

Combining equations (12), (13) and (14), the following equation is obtained:

$$\left(\frac{d\sqrt{I}}{dV_G}\right)^2 = \frac{W}{4L}\frac{\left[\sigma(V_G) + \sum_{i=1}^{\infty}\frac{\sigma^{(i)}(V_G)}{(i)!}(-V_{S^*})^i - \sum_{i=0}^{\infty}\frac{\sigma^{(i)}(V_G)}{(i)!}(-V_{S^*})^i\cdot\frac{dV_{S^*}}{dV_G}\right]^2}{\left|-\sum_{i=0}^{\infty}\frac{\sigma^{(i)}(V_G)}{(i+1)!}\left((-V_G)^{i+1} - (-V_{S^*})^{i+1}\right)\right|} \tag{15}$$

In order to write equation (15) in a more useful form, it is necessary to perform some simplifications. First it is assumed that $\frac{dV_{S^*}}{dV_G} \ll 1$ so that we can remove it from the numerator of the fraction. Furthermore, keeping in mind that the conductivity of the depleted channel is negligible, i.e. $\sigma(0) = 0$, it is possible to write:

$$\sigma(0) = \sum_{i=0}^{\infty}\frac{\sigma^{(i)}(V_G)}{i!}(-V_G)^i \Rightarrow 0 = \sigma(V_G) + \sum_{i=1}^{\infty}\frac{\sigma^{(i)}(V_G)}{i!}(-V_G)^i \Rightarrow \sigma(V_G) = -\sum_{i=1}^{\infty}\frac{\sigma^{(i)}(V_G)}{i!}(-V_G)^i \tag{16}$$

Then equation (12) can be written as:

$$\left(\frac{d\sqrt{I}}{dV_G}\right)^2 = \frac{W}{4L}\frac{\left[-\sum_{i=1}^{\infty}\frac{\sigma^{(i)}(V_G)}{i!}(-V_G)^i + \sum_{i=1}^{\infty}\frac{\sigma^{(i)}(V_G)}{i!}(-V_{S^*})^i\right]^2}{\left|-\sum_{i=0}^{\infty}\frac{\sigma^{(i)}(V_G)}{(i+1)!}\left((-V_G)^{i+1} - (-V_{S^*})^{i+1}\right)\right|}$$

$$= \frac{W}{4L}\frac{\left[-\sum_{i=1}^{\infty}\frac{\sigma^{(i)}(V_G)}{i!}(-V_G)^i + \sum_{i=1}^{\infty}\frac{\sigma^{(i)}(V_G)}{i!}(-V_{S^*})^i\right]^2}{\left|\sigma(V_G)V_{GS^*} - \sum_{i=1}^{\infty}\frac{\sigma^{(i)}(V_G)}{(i+1)!}\left((-V_G)^{i+1} - (-V_{S^*})^{i+1}\right)\right|} \tag{17}$$

$$= \frac{W}{4L}\frac{\left[-\sum_{i=1}^{\infty}\frac{\sigma^{(i)}(V_G)}{i!}(-V_G)^i + \sum_{i=1}^{\infty}\frac{\sigma^{(i)}(V_G)}{i!}(-V_{S^*})^i\right]^2}{\left|-\sum_{i=1}^{\infty}\frac{\sigma^{(i)}(V_G)}{i!}(-V_G)^i V_{GS^*} - \sum_{i=1}^{\infty}\frac{\sigma^{(i)}(V_G)}{(i+1)!}\left((-V_G)^{i+1} - (-V_{S^*})^{i+1}\right)\right|}$$

By separating the contribution of the first derivative and including the rest of the terms of the Taylor series expansion in a parameter $k(V_G)$, it is possible to rewrite equation (17) as:

$$\left(\frac{d\sqrt{I}}{dV_G}\right)^2 = \frac{W}{4L} \frac{\sigma'(V_G)^2 V_{GS^*}^2}{\left|\sigma'(V_G)V_{GS^*}V_G - \frac{1}{2}\sigma'(V_G)((V_G)^2 - (V_{S^*})^2)\right|} \cdot k(V_G)$$
$$= \frac{W}{4L} \frac{\sigma'(V_G)}{\left|\frac{V_G}{V_{GS^*}} - \frac{1}{2}\left(\frac{V_G + V_{S^*}}{V_{GS^*}}\right)\right|} = \frac{W}{4L} \frac{\sigma'(V_G)}{\left|\frac{1}{2}\left(\frac{V_G - V_{S^*}}{V_{GS^*}}\right)\right|} = \frac{W}{2L}\sigma'(V_G) \cdot k(V_G) \tag{18}$$

where $k(V_G)$ depends on the particular kind of charge density dependence of the mobility. In the simpler case of a power law dependence of the mobility:

$$\mu(V)C(V) = C_i \mu_0 (V - V_{th})^\beta \tag{19}$$

it can be easily seen how equation (18) is reduced to:

$$\left(\frac{d\sqrt{I}}{dV_G}\right)^2 = \frac{W}{2L}\mu(V_G)C(V_G) \cdot k(V_G), \quad k = \frac{\beta+2}{2(\beta+1)} \tag{20}$$

and the value of $k$ becomes gate voltage independent (for the full derivation see below). Most importantly, it can be seen that:

$$0 \leq \beta \leq +\infty \quad \Rightarrow \quad 1 \geq k \geq 0.5$$
$$-1 < \beta < 0 \quad \Rightarrow \quad +\infty > k > 1 \tag{21}$$

In the physically relevant of a mobility increasing with charge density, the correction factor $k(V_G)$ will therefore be limited between 1 and 0.5 although it will not be constant throughout the curve. If it is possible to fit the function $\left(\frac{d\sqrt{I}}{dV_G}\right)^2$ with the power law dependence in the range used for the measurement, the extracted value of $\beta$ can then be used to approximate the value of $k$ to a constant value, and the mobility in the saturation regime can be obtained as:

$$\mu(V_G) = \frac{2L}{WC(V_G) \cdot k}\left(\frac{d\sqrt{I}}{dV_G}\right)^2 \tag{22}$$

without the need to perform a second derivative of the transfer characteristics. This approximation holds also when the approximation with a power law dependence is not perfect due to the relatively little dependence of the value of $k$ with respect to the value of $\beta$. The value of $C(V_G)$ can be easily obtained either by calculating the geometric capacitance of the dielectric layer or by measuring the capacitance-voltage characteristics of the actual device.

The comparison of corrected and uncorrected mobility with the second derivative method and the experimental methods is shown in figure S1. The same approach is followed in the main text, showing in that case an even better agreement with the four point probe measurements.

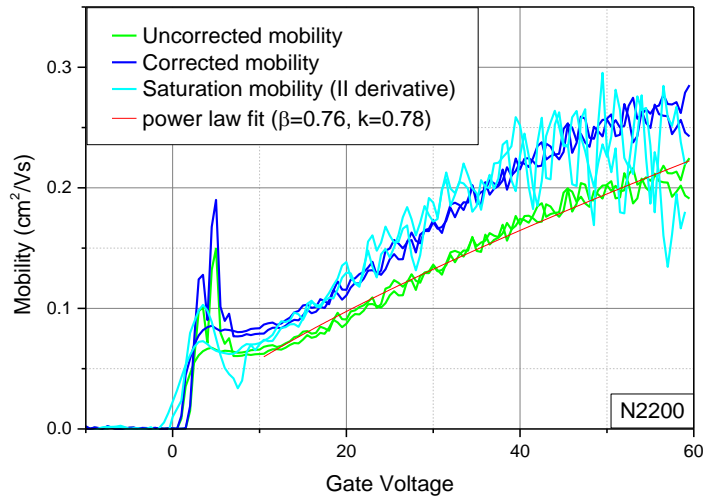

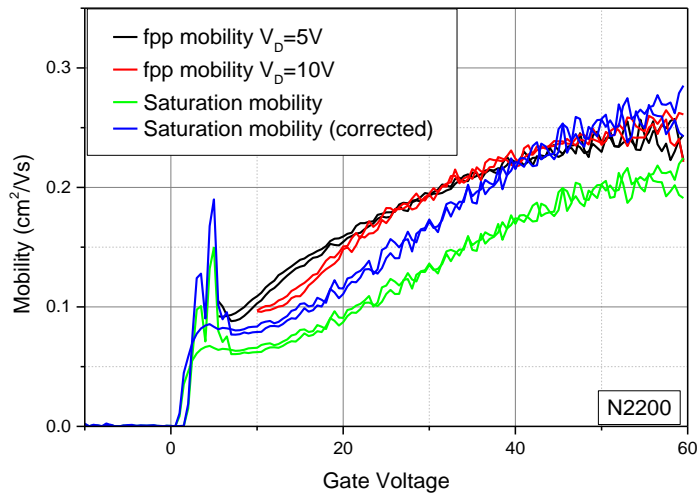

Figure S1. (a) Comparison of the different mobility extraction methods, the uncorrected mobility (*k*=1), the corrected mobility (equation (22)) obtained with a power law approximation (red trace) and the second derivative method (equation (12)) with a 15 point Savitzky-Golay filtering (third order polynomial). (b) Four point probe mobility (for $V_{DS}$=5V and $V_{DS}$=10V) and the corrected and uncorrected mobility showing the higher agreement with the correction.

# 3 - Full derivation of the saturation mobility equation for a power law dependent mobility

Starting from the following function:

$$\mu(V)C(V) = C_i\mu_0(V - V_{th})^\beta \tag{23}$$

It is possible to explicitly calculate the current-voltage characteristics of the device using the following equation (the inner integral lower limit needs to be shifted to $V_{th}$, as function (23) is only defined for $V_G \geq V_{th}$ and is identically 0 below that value):

$$IL = W \int_0^{V_G} \int_{V_{th}}^{V_G-V(x)} C_i\mu_0(v - V_{th})^\beta dv\, dV(x) \tag{24}$$

Keeping $V_G$ as a parameter, the double integration can be performed by taking into account that the generalized integral is zero for lower integration parameter:

$$\int_0^{V_G} \int_{V_{th}}^{V_G-V(x)} C_i\mu_0(v - V_{th})^\beta dv\, dV(x) = \int_0^{V_G-V_{th}} \frac{1}{\beta+1} C_i\mu_0(V_G - V(x) - V_{th})^{\beta+1} dV(x) =$$

$$\frac{1}{(\beta+1)(\beta+2)} C_i\mu_0(V_G - V_{th})^{\beta+2} = \frac{1}{(\beta+1)(\beta+2)} C_i\mu_0(V_G - V_{th})^{\beta+2} \tag{25}$$

Where the limit on the outer integral is shifted always to keep into account the presence of the threshold voltage (channel is pinched-off, that is charge density at the upper voltage limit is 0, if threshold voltage is not taken into account it would become negative)

Therefore the current can be written as:

$$I = \frac{1}{(\beta+1)(\beta+2)} \frac{W}{L} C_i\mu_0(V_G - V_{th})^{\beta+2} \tag{26}$$

And the saturation mobility can be extracted from the following derivation:

$$\left(\frac{d\sqrt{I}}{dV_G}\right)^2 = \frac{W}{L} \frac{1}{(\beta+1)(\beta+2)} C_i\mu_0 \left(\frac{d(V_G-V_{th})^{\frac{\beta+2}{2}}}{dV_G}\right)^2 = \frac{W}{L} \frac{(\beta+2)^2}{4(\beta+1)(\beta+2)} C_i\mu_0 \left((V_G - V_{th})^{\frac{\beta}{2}}\right)^2 = \frac{W}{2L} C_i\mu_0(V_G - V_{th})^\beta \frac{(\beta+2)}{2(\beta+1)} = \frac{W}{2L} \mu(V_G)C(V_G) \frac{(\beta+2)}{2(\beta+1)} \tag{27}$$

With a correction factor now gate voltage independent and equal to $(V_G) = \frac{(\beta+2)}{2(\beta+1)}$.

As a final remark, a more accurate expression of the power law mobility function would require the introduction of a theta function to ensure that the function is identically 0 below threshold. That would not require the artificial shifting of the integration boundaries, however the result would not be different.

## 4 - Example of correction factor approximation

As an example of how the correction factor can be estimated for an arbitrary function, the transfer characteristics have been simulated starting from the following polynomial function:

$$\mu(V)C(V) = 1.17 \cdot \theta(V) \cdot (0.1 \cdot V^{0.5} + 2 \cdot 10^{-5} \cdot V^{2.5})$$

In figure S2(a) the chosen mobility function is shown alongside the two power law components, responsible for the low voltage and high voltage part of the plot. A simple power law fit is shown in a dashed line.

Using equation (22) mobility is extracted from the calculated transfer characteristics (figure S2(b)), without any correction factor (in red) and using the approximated correction factor obtained from the power law fit (in blue). The procedure returns a much more accurate mobility value despite the rather poor agreement between the original curve and the fit.

The reason for this is shown in figure S2(c), where the mobility ratio (equal to $1/k$) is calculated for the full curve (black trace), for the two power components of equation (23) (red and green traces) and for the power law fit of the chosen function. As shown in the mathematical derivation, the correction factor for an arbitrary function is in general gate voltage dependent, but in any case is within the values of the exponents of the different monomers which are used in the polynomial function. The rough approximation resulting from the power law fit returns a value that is an average of the complete gate voltage dependent correction factor. Although it does not fully recover the initial mobility, it greatly improves the accuracy of the mobility extraction, which is in turn fundamental for the contact resistance extraction.

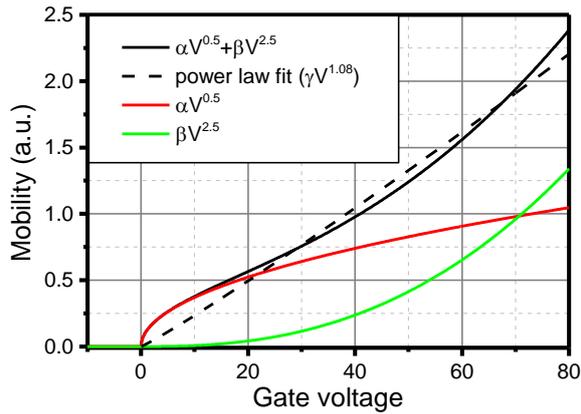

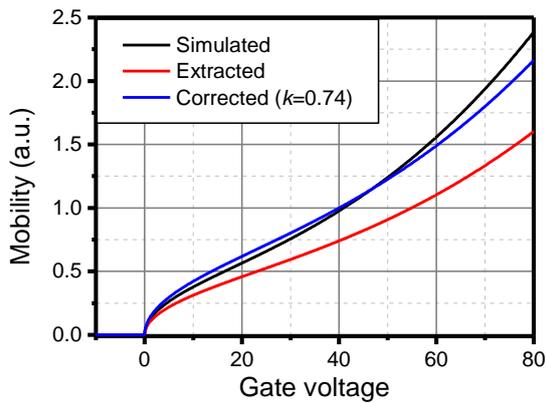

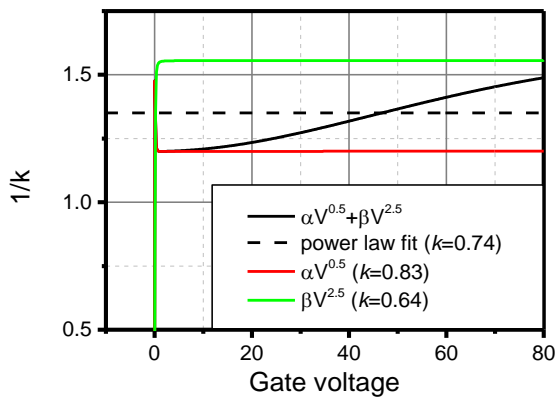

Figure S2. (a) A model polynomial mobility curve, used for the calculation of the transfer characteristics (black trace) is shown together with the two power law components (red and green trace) and a power law fit (dashed trace). (b) The model mobility curve (black trace), together with the uncorrected mobility extracted from the calculated transfer characteristics (red trace) and the saturation mobility corrected using the scaling factor extracted from the power law fit (blue trace). (c) Calculated correction factors for the full polynomial mobility (black trace, gate voltage dependent), for the two different components of the polynomial mobility (red and green) demonstrating how the real correction factor is always in between the two extreme values, and the correction factor obtained from the power law fit (averaging the variable correction factor of the initial function.

## 5 - Extended experimental details

All the devices were prepared in a bottom-contact top-gate geometry. A hall bar structure with patterned semiconductor [1] has been used for the four point probe measurement (W=200µm, L=100µm, probe at ¼ and ¾ of the channel length), while a pre-patterned substrate with different channel length transistors (W=2.85mm, L=5/12/50µm) has been used for the TLM measurements. Devices for the OSC thickness dependent measurement have been performed on interdigitated source-drain electrodes prepared by shadow mask evaporation (W=149mm, L=100µm, contact thickness 20nm). The effective contact length, required for the extraction of contact resistivity for the interdigitated pattern is defined as half the width of the contact finger ($L_C = 50 \ \mu m$). P(NDI2OD-T2) and PBTTT, an n- and a p-type organic semiconductor respectively, were dissolved in dichlorobenzene for the four point probe devices, while P(NDI2OD-T2) dissolved in a 1:1 mixture of chloronaphthalene:xylene was used for the TLM devices. All solutions had an 8 g/l concentration (except for the variable thickness measurements where different concentrations from 5 to 20 g/l were used) and were deposited via spin-coating at 1000 rpm for 60 seconds and annealed at 200°C and 180°C for P(NDI2OD-T2) and PBTTT, respectively. A dielectric layer made of EG-PMMA (Polymer Source P8509P-MMA) with a concentration of 60 g/l was subsequently spin-cast at 1500 rpm for 90 seconds and dried overnight at 80°C. Aluminum gate electrodes were deposited through shadow mask evaporation. All measurements have been performed at room temperature in inert atmosphere using an Agilent 4155C semiconductor parameter analyzer and a needle probe station, in all cases performing forward and backwards voltage sweeps. The capacitance was determined from the measured thickness of the PMMA layer (roughly 550 nm for all devices). All thickness measurements were performed using a Dektak³ST stylus profilometer.

## 6 - N2200 four-point probe mobility extraction

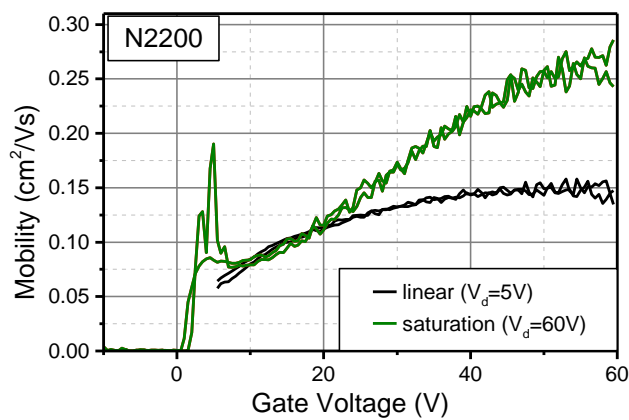

Figure S3. Extracted mobility from the linear and saturation transfer curves of an N2200 four point probe transistor, using equation (9) and (18) respectively.

Figure S3 shows the extracted mobility of an N2200 transistor. At low gate voltages the saturation mobility becomes lower than the linear one, thus preventing the calculation of any contact resistance. The unexpected behavior is possibly due to the additional preparation steps required for pattering the semiconductor, as the same effect is not present in any other N2200 device. This could lead to an invalidation of the assumption $\frac{dV_{S^*}}{dV_G} \ll 1$, necessary for the extraction of accurate values of the mobility in the saturation regime. However, at higher gate voltages the correct dependence is recovered and the model can be used to correctly estimate the contact resistance (as reported in the main text).

## 7 - Uncertainty estimations

The measurement accuracy for the used equipment (Agilent 4155C) is dependent on many factors including measurement integration time, applied voltage and measurement range. For the voltages and ranges of interest in this work, it's safe to assume that the relative error on a current measurement is constant and equal to $\xi_i = 10^{-3}$.

The numerical derivation for a generic data point *n* is performed according to the following formula:

$$\gamma_n = \left.\frac{dI_D}{dV_G}\right|_n = \frac{1}{2}\left(\frac{I_{D_n} - I_{D_{n-1}}}{V_{G_n} - V_{G_{n-1}}} + \frac{I_{D_{n+1}} - I_{D_n}}{V_{G_{n+1}} - V_{G_n}}\right)$$

With the relative error that can be calculated as:

$$\frac{\xi_{\gamma_n}}{\gamma_n} \cong \frac{\xi_i \cdot I_{D_n}}{I_{D_n} - I_{D_{n-1}}}$$

And for the saturation regime it is possible to derive an analogous formula:

$$\frac{\xi_{\varphi_n}}{\varphi_n} \cong \frac{\xi_i \cdot I_{D_n}}{2(I_{D_n} - I_{D_{n-1}})}$$

with $\varphi_n = \left.\frac{d\sqrt{I_D}}{dV_G}\right|_n$.

Starting from the equations above, it is possible to determine the uncertainty for mobility and contact resistance using a simple error propagation rule for uncorrelated values:

$$\left.\frac{\xi_{\mu_{sat}}}{\mu_{sat}}\right|_n = \left.\frac{\xi_{\mu_{lin}}}{\mu_{sat}}\right|_n \cong \frac{\xi_i \cdot I_{D_n}}{I_{D_n} - I_{D_{n-1}}}$$

$$\left.\frac{\xi_{R_C}}{R_C}\right|_n \cong \left.\frac{\xi_{V_{D'S'}}}{V_{D'S'}}\right|_n \cong \sqrt{\left(\frac{\xi_i \cdot I_{D\,n}^{sat}}{I_{D\,n}^{sat} - I_{D\,n-1}^{sat}}\right)^2 + \left(\frac{\xi_i \cdot I_{D\,n}^{lin}}{I_{D\,n}^{lin} - I_{D\,n-1}^{lin}}\right)^2}$$

The result of this analysis is shown in figure S4 for the four point probe measurement, where the difference between the measured and calculated values (relative to the calculated values) is compared against the accuracy of the calculated values (taken as twice the standard deviation).

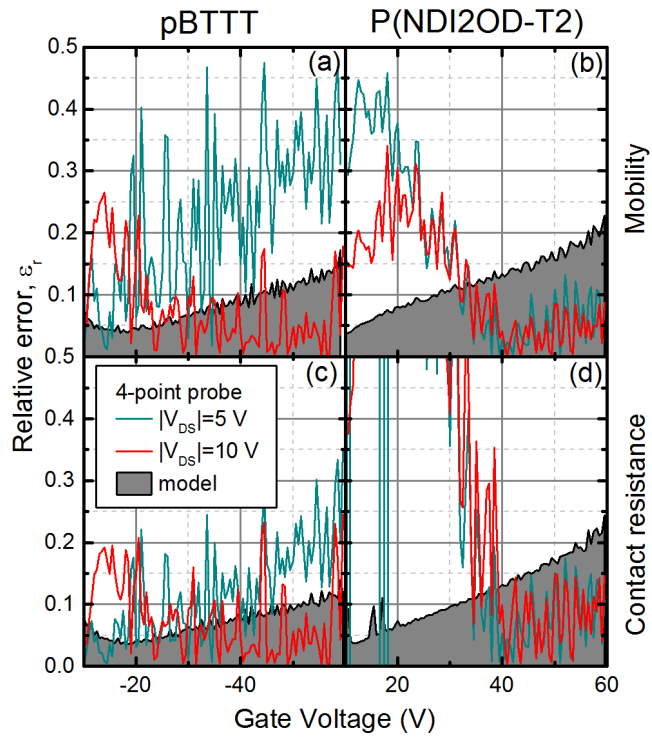

Figure S4. The difference between the calculated and experimental values reported in figure 2 (main article) relative to the calculated values is plotted: (a) pBTTT mobility, (b) P(NDI2OD-T2) mobility, (c) pBTTT contact resistance, (d) P(NDI2OD-T2) contact resistance. These are compared against the accuracy of the calculated values, displayed as the black lines. Values in the gray shaded area are not significant since differences are below the accuracy of the calculated values. Only the calculation on the backwards traces are shown for clarity.

## 8 - N2200 four-point probe mobility extraction

Figure S5 shows the total device resistance at different source-gate voltages, obtained by fitting the output curve between $V_{DS} = 1V$ and $V_{DS} = 4V$.

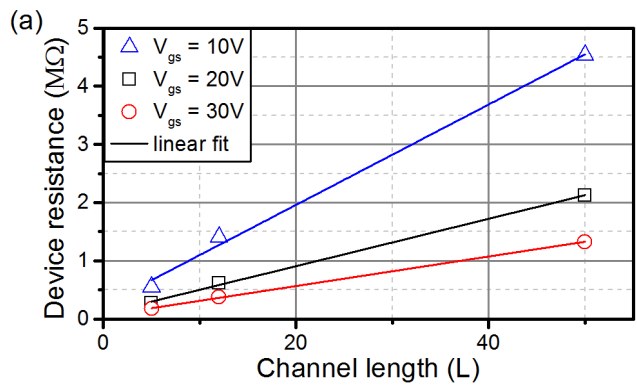

Figure S5. Total device resistance of the TLM devices plotted against channel length, for different gate voltages. The linear fits used for the calculation of contact and channel resistance are also shown in the plot for each data set.

[1]    J.-F. Chang, M. C. Gwinner, M. Caironi, T. Sakanoue, and H. Sirringhaus, Advanced Functional Materials **20**, 2825 (2010).